\documentclass[nofootinbib]{revtex4}%
\usepackage{hyperref}
\usepackage{amsmath,amssymb}
\usepackage{graphicx}
\usepackage{epstopdf}
\usepackage{amsmath}
\usepackage{amsfonts}
\usepackage{amssymb}%
\providecommand{\U}[1]{\protect\rule{.1in}{.1in}}
\begin{document}
\title{Phase Structure in a Dynamical Soft-Wall Holographic QCD Model}
\author{Song He}
\email{hesong@itp.ac.cn}
\affiliation{State Key Laboratory of Theoretical Physics, Institute of Theoretical Physics,
Chinese Academy of Science, Beijing 100190, PRC}
\author{Shang-Yu Wu}
\email{loganwu@gmail.com}
\affiliation{Institute of Physics, National Chiao Tung University, Hsinchu, ROC}
\author{Yi Yang}
\email{yiyang@mail.nctu.edu.tw}
\affiliation{Department of Electrophysics, National Chiao Tung University}
\affiliation{National Center for Theoretical Science, Hsinchu, ROC}
\author{Pei-Hung Yuan}
\email{phy.pro.phy@gmail.com}
\affiliation{Institute of Physics, National Chiao Tung University, Hsinchu, ROC}
\date{\today}

\begin{abstract}
We consider the Einstein-Maxwell-dilaton system with an arbitrary kinetic
gauge function and a dilaton potential. A family of analytic solutions is
obtained by the potential reconstruction method. We then study its holographic
dual QCD model. The kinetic gauge function can be fixed by requesting the
linear Regge spectrum of mesons. We calculate the free energy to obtain
the phase diagram of the holographic QCD model and interpret our result as the
heavy quarks system by comparing the recent lattice QCD simulation. We finally
obtain the equations of state in our model.

\end{abstract}
\maketitle
\tableofcontents
%

\setcounter{equation}{0}
\renewcommand{\theequation}{\arabic{section}.\arabic{equation}}%

\section{Introduction}

To study phase structure of QCD is a challenging and important task. It is well
known that QCD is in the confinement and chiral symmetry breaking phase for the low temperature and small chemical potential, while it is in the
deconfinement and chiral symmetry restored phase for the high temperature and
large chemical potential. Thus it is widely believed that there exists a phase
transition between these two phases. To obtain the phase transition line in
the $T-\mu$ phase diagram is a rather difficult task because the QCD coupling
constant becomes very strong near the phase transition region and the
conventional perturbative method does not work well. Moreover, with the
nonzero physical quark masses presented, part of the phase transition line
will weaken to a crossover for a range of temperature and chemical potential
that makes the phase structure of QCD more complicated. Locating the critical
point where the phase transition converts to a crossover is another challenging work. For a long time, lattice QCD is the only method to attack these problems.
Although lattice QCD works well for zero density, it encounters the sign
problem when considering finite density, i.e. $\mu\neq0$. However, the most
interesting region in the QCD phase diagram is at finite density. The most
concerned subjects, such as heavy-ion collisions and compact stars in
astrophysics, are all related to QCD at finite density. Recently, lattice QCD
has developed some techniques to solve the sign problem, such as reweighting
method, imaginary chemical potential method and the method of expansion in
$\mu/T$. Nevertheless, these techniques are only able to deal with the cases
of small chemical potentials and quickly lost control for the larger chemical
potential. See \cite{1009.4089} for a review of the current status of lattice QCD.

On the other hand, using the idea of AdS/CFT duality from string theory, one
is able to study QCD in the strongly coupled region by studying its weakly
coupled dual gravitational theory, i.e. holographic QCD. The models which are
directly constructed from string theory are called the top-down models. The
most popular top-down models are D3-D7 \cite{0306018,0311270,0304032,0611099} model
and D4-D8 (Sakai-Sugimoto) model \cite{0412141,0507073}. In these top-down
holographic QCD models, confinement and chiral symmetry phase transitions in
QCD have been addressed and been translated into geometric transformations in the
dual gravity theory. Meson spectrums and decay constants have also been
calculated and compared with the experimental data with surprisingly consistency.
Although the top-down QCD models describe many important properties in
realistic QCD, the meson spectrums obtained from those models can not realize
the linear Regge trajectories. To solve this problem, another type of
holographic models have been developed, i.e. bottom-up models, such as the
hard wall model \cite{0501128} and the later refined soft-wall model
\cite{0602229}. In the original soft-wall model, the IR correction of the
dilaton field was put by hand to obtain the linear Regge behavior of the meson
spectrum. However, since the fields configuration is put by hand, it does not
satisfy the equations of motion. To get a fields configuration which is both
consistent with the equation of motions and realizes the linear Regge
trajectory, dynamical soft-wall models were constructed by introduce a dilaton
potential \cite{0801.4383,0806.3830} consistently. Later on, the Einstein-dilaton and
Einstein-Maxwell-dilaton models have been widely studied numerically
\cite{0804.0434,1006.5461,1012.1864,1108.2029,1201.0820}. By potential reconstruction method,
analytic solutions can be obtained in the Einstein-dilaton model
\cite{1103.5389} and similarly in the Einstein-Maxwell-dilaton model
\cite{1201.0820,1209.4512}.

In this paper, we consider the Einstein-Maxwell-dilaton system with an
arbitrary kinetic gauge function and a dilaton potential. A family of analytic
solutions are obtained by the potential reconstruction method. We then study
its holographic dual QCD model. The kinetic gauge function can be fixed by
requesting the meson spectrums satisfy the linear Regge trajectories. We
study the thermodynamics of the Einstein-Maxwell-dilaton background and
calculate the free energy to obtain the phase diagram of the holographic QCD
model. By comparing our result with the recent lattice QCD simulations, we interpret our system as the model for the heavy quarks system. We
finally compute the different equation of states in our model and discuss
their behaviors. The behavior of the equations of state is consistent with our
interpretation of the heavy quarks.

The paper is organized as follows. In section II, we consider the
Einstein-Maxwell-dilaton system with a dilaton potential as well as a gauge
kinetic function. By potential reconstruction method, we obtain a family of
analytic solutions with arbitrary gauge kinetic function and warped factor. We
then fix the gauge kinetic function by requesting the meson spectrums to
realize the linear Regge trajectories. By choosing a proper warped factor, we
obtain the final form of our analytic solution. In section III, we study the
thermodynamics of our gravitational background and compute the free energy to
get the phase diagram. From the phase diagram, we argue that our background is
dual to QCD with heavy quarks and interpret the black hole phase
transition as the deconfinement phase transition in QCD. We finally plot the
equations of state in our background to compare with that in QCD. We conclude
our result in section IV.

%

\setcounter{equation}{0}
\renewcommand{\theequation}{\arabic{section}.\arabic{equation}}%

\section{Einstein-Maxwell-Dilaton System}

We consider a 5-dimensional Einstein-Maxwell-dilaton system with probe matters.
The action of the system have two parts, the background part and the matter
part,%
\begin{equation}
S=S_{b}+S_{m}.
\end{equation}
In string frame, the background action includes a gravity field $g_{\mu\nu
}^{s}$, a Maxwell field $A_{\mu}$ and a neutral dilatonic scalar field
$\phi_{s}$,%
\begin{equation}
S_{b}=\dfrac{1}{16\pi G_{5}}\int d^{5}x\sqrt{-g^{s}}e^{-2\phi_{s}}\left[
{R}_{s}-{\frac{f_{s}\left(  \phi_{s}\right)  }{4}F^{2}}+4\partial_{\mu}%
\phi_{s}\partial^{\mu}\phi_{s}-V_{s}\left(  \phi_{s}\right)  \right]  ,
\label{background}%
\end{equation}
where $f_{s}\left(  \phi_{s}\right)  $ is the gauge kinetic function
associated to the Maxwell field $A_{\mu}$ and $V_{s}\left(  \phi_{s}\right)  $
is the potential of the dilaton field. One should note that the function
$f_{s}\left(  \phi_{s}\right)  $ is a positive-definite function. The explicit
forms of the gauge kinetic function $f_{s}\left(  \phi_{s}\right)  $ and the
dilaton potential $V_{s}\left(  \phi_{s}\right)  $ are not given ad hoc and
will be solved consistently with the background.

The matter action includes the flavor fields $\left(  {A}_{\mu}^{L},{A}_{\mu
}^{R}\right)  $, which we will treat as probe, describing the degree of
freedom of mesons on the 4d boundary,%
\begin{equation}
S_{m}=-\dfrac{1}{16\pi G_{5}}\int d^{5}x\sqrt{-g^{s}}e^{-2\phi_{s}}%
{\frac{f_{s}\left(  \phi_{s}\right)  }{4}}\left(  {F_{L}^{2}+F_{R}^{2}%
}\right)  , \label{matter}%
\end{equation}
where ${G}_{5}$ is the coupling constant for the flavor field strength
${F}_{\mu\nu}{=\partial}_{\mu}A_{\nu}-{\partial}_{\nu}A_{\mu}$ and
$f_{s}\left(  \phi_{s}\right)  $ is the gauge kinetic function associated to
the flavor field. The gauge kinetic function of the flavor field in the matter
action is not necessary to be the same as that of the Maxwell field in the
background action [\ref{background}] in general, but here we set them equal
for simplicity.

In the above, we have defined our model in string frame in which it is natural
to write the boundary conditions as we will see when solving the background in
the next section. However, to study the thermodynamics of QCD at finite
temperature, it is convenient to solve the equations of motion and study the
equations of state in Einstein frame. To transform the action from string
frame to Einstein frame, we make the following "standard" transformations,%
\begin{equation}
\phi_{s}=\sqrt{\dfrac{3}{8}}\phi\text{, \ \ }g_{\mu\nu}^{s}=g_{\mu\nu}%
{e}^{\sqrt{\tfrac{2}{3}}\phi}\text{, \ \ }f_{s}\left(  \phi_{s}\right)
=f\left(  \phi\right)  {e}^{\sqrt{\tfrac{2}{3}}\phi}\text{, \ \ }V_{s}\left(
\phi_{s}\right)  ={e}^{-\sqrt{\tfrac{2}{3}}\phi}V\left(  \phi\right)  .
\end{equation}
The background and the matter actions become, in Einstein frame,%
\begin{align}
S_{b}  &  =\dfrac{1}{16\pi G_{5}}\int d^{5}x\sqrt{-g}\left[  {R-\frac{f\left(
\phi\right)  }{4}F^{2}}-\dfrac{1}{2}\partial_{\mu}\phi\partial^{\mu}%
\phi-V\left(  \phi\right)  \right]  ,\label{action-b}\\
S_{m}  &  =-\dfrac{1}{16\pi G_{5}}\int d^{5}x\sqrt{-g}{\frac{f\left(
\phi\right)  }{4}}\left(  {F_{V}^{2}+F_{\tilde{V}}^{2}}\right)  .
\label{action-m}%
\end{align}
where we have written the flavor fields\ ${A}^{L}$ and ${A}^{R}$\ in terms of
the vector meson and pseudovector meson fields $V$ and $\tilde{V}$,%
\begin{equation}
{A}^{L}=V+\tilde{V}\text{, \ \ }{A}^{R}=V-\tilde{V}.
\end{equation}
The equations of motion can be derived from the actions (\ref{action-b}) and
(\ref{action-m}) as%
\begin{align}
\nabla^{2}\phi &  =\frac{\partial V}{\partial\phi}+\frac{1}{4}\frac{\partial
f}{\partial\phi}\left(  F^{2}+{F_{V}^{2}+F_{\tilde{V}}^{2}}\right)
,\label{eom1}\\
\nabla_{\mu}\left[  f(\phi)F^{\mu\nu}\right]   &  ={{0,}}\\
\nabla_{\mu}\left[  f(\phi)F_{V}^{\mu\nu}\right]   &  ={{0,}}\label{eom3}\\
\nabla_{\mu}\left[  f(\phi)F_{\tilde{V}}^{\mu\nu}\right]   &  ={0,}\\
R_{\mu\nu}-\frac{1}{2}g_{\mu\nu}R  &  =\frac{f(\phi)}{2}\left(  F_{\mu\rho
}F_{\nu}^{\rho}-\frac{1}{4}g_{\mu\nu}F^{2}+\left\{  {F_{V},F_{\tilde{V}}%
}\right\}  \right)  +\frac{1}{2}\left[  \partial_{\mu}\phi\partial_{\nu}%
\phi-\frac{1}{2}g_{\mu\nu}\left(  \partial\phi\right)  ^{2}-g_{\mu\nu
}V\right]  . \label{eom5}%
\end{align}
In the next section, we will solve the above equations of motion under some
physical boundary conditions and constraints.

\subsection{The gravitational Background}

In this section, we will solve the background of the Einstein-Maxwell-dilaton
system defined in the last section. We first turn off the probe flavor field
$V$ and $\tilde{V}$ in the equations of motion (\ref{eom1}-\ref{eom5}), which
reduce to%
\begin{align}
\nabla^{2}\phi &  =\frac{\partial V}{\partial\phi}+\frac{F^{2}}{4}%
\frac{\partial f}{\partial\phi},\\
\nabla_{\mu}\left[  f(\phi)F^{\mu\nu}\right]   &  ={{0,}}\\
R_{\mu\nu}-\frac{1}{2}g_{\mu\nu}R  &  =\frac{f(\phi)}{2}\left(  F_{\mu\rho
}F_{\nu}^{\rho}-\frac{1}{4}g_{\mu\nu}F^{2}\right)  +\frac{1}{2}\left[
\partial_{\mu}\phi\partial_{\nu}\phi-\frac{1}{2}g_{\mu\nu}\left(  \partial
\phi\right)  ^{2}-g_{\mu\nu}V\right]  .
\end{align}
Because we are interested in the black\ hole solutions, we consider the
following form of the metric in Einstein frame,%
\begin{align}
ds^{2}  &  =\dfrac{L^{2}e^{2A\left(  z\right)  }}{z^{2}}\left[  -g(z)dt^{2}%
+\frac{dz^{2}}{g(z)}+d\vec{x}^{2}\right]  ,\label{metric}\\
\phi &  =\phi\left(  z\right)  \text{, \ \ }A_{\mu}=A_{t}\left(  z\right),
\label{ansatz}%
\end{align}
where $z=0$ corresponds to the conformal boundary of the 5d spacetime. We will
set the radial $L$ of $AdS_{5}$ space to be unit in the following of this paper.

Using the ansatz of the metric, the Maxwell field and the dilaton field
(\ref{metric}, \ref{ansatz}), the equations of motion and constraints for the
background fields become%
\begin{align}
\phi^{\prime\prime}+\left(  \frac{g^{\prime}}{g}+3A^{\prime}-\dfrac{3}%
{z}\right)  \phi^{\prime}+\left(  \frac{z^{2}e^{-2A}A_{t}^{\prime2}f_{\phi}%
}{2g}-\frac{e^{2A}V_{\phi}}{z^{2}g}\right)   &  =0,\label{eom-phi}\\
A_{t}^{\prime\prime}+\left(  \frac{f^{\prime}}{f}+A^{\prime}-\dfrac{1}%
{z}\right)  A_{t}^{\prime}  &  =0,\label{eom-At}\\
A^{\prime\prime}-A^{\prime2}+\dfrac{2}{z}A^{\prime}+\dfrac{\phi^{\prime2}}{6}
&  =0,\label{eom-A}\\
g^{\prime\prime}+\left(  3A^{\prime}-\dfrac{3}{z}\right)  g^{\prime}%
-e^{-2A}z^{2}fA_{t}^{\prime2}  &  =0,\label{eom-g}\\
A^{\prime\prime}+3A^{\prime2}+\left(  \dfrac{3g^{\prime}}{2g}-\dfrac{6}%
{z}\right)  A^{\prime}-\dfrac{1}{z}\left(  \dfrac{3g^{\prime}}{2g}-\dfrac
{4}{z}\right)  +\dfrac{g^{\prime\prime}}{6g}+\frac{e^{2A}V}{3z^{2}g}  &  =0.
\label{eom-V}%
\end{align}
We should notice that only four of the above five equations are
independent. In the following, we will solve the equations (\ref{eom-At}%
-\ref{eom-V}), and leave the equation (\ref{eom-phi}) as a constraint for a
consistent check.

To solve the background, we need to specify the boundary conditions. Near the
horizon $z=z_{H}$, we require%
\begin{equation}
A_{t}\left(  z_{H}\right)  =g\left(  z_{H}\right)  =0, \label{bc-zH}%
\end{equation}
due to the physical requirement that $A_{\mu}A^{\mu}=g^{tt}A_{0}A_{0}$ must be
finite at $z=z_{H}$.

Near the boundary $z\rightarrow0$, we require the metric in string frame
to be asymptotic to $AdS_{5}$, thus%
\begin{equation}
ds_{z\rightarrow0}^{2}=g_{\mu\nu}^{s}\left(  z\rightarrow0\right)  dx^{\mu
}dx^{\nu}=\dfrac{1}{z^{2}}\left[  -dt^{2}+dz^{2}+d\vec{x}^{2}\right]  .
\end{equation}
This boundary condition, in Einstein frame, becomes,%
\begin{equation}
A\left(  0\right)  =-\sqrt{\dfrac{1}{6}}\phi\left(  0\right)  \text{,
\ \ }g\left(  0\right)  =1. \label{bc-0}%
\end{equation}
Since we do not assume the form of the dilaton potential $V\left(
\phi\right)  $, which should be solved consistently from the equations of
motion, we will treat the dilaton potential as a function of $z$, i.e.
$V\left(  z\right)  $, when we solve the equations of motion. With the above
boundary conditions (\ref{bc-zH}) and (\ref{bc-0}), the equations of motion
(\ref{eom-At}-\ref{eom-V}) can be analytically solved as%
\begin{align}
\phi^{\prime}\left(  z\right)   &  =\sqrt{-6\left(  A^{\prime\prime}%
-A^{\prime2}+\dfrac{2}{z}A^{\prime}\right)  },\label{phip}\\
A_{t}\left(  z\right)   &  =\sqrt{\dfrac{-1}{\int_{0}^{z_{H}}y^{3}%
e^{-3A}dy\int_{y_{g}}^{y}\dfrac{x}{e^{A}f}dx}}\int_{z_{H}}^{z}\dfrac{y}%
{e^{A}f}dy,\label{At}\\
g\left(  z\right)   &  =1-\dfrac{\int_{0}^{z}y^{3}e^{-3A}dy\int_{y_{g}}%
^{y}\dfrac{x}{e^{A}f}dx}{\int_{0}^{z_{H}}y^{3}e^{-3A}dy\int_{y_{g}}^{y}%
\dfrac{x}{e^{A}f}dx},\label{gp}\\
V\left(  z\right)   &  =-3z^{2}ge^{-2A}\left[  A^{\prime\prime}+3A^{\prime
2}+\left(  \dfrac{3g^{\prime}}{2g}-\dfrac{6}{z}\right)  A^{\prime}-\dfrac
{1}{z}\left(  \dfrac{3g^{\prime}}{2g}-\dfrac{4}{z}\right)  +\dfrac
{g^{\prime\prime}}{6g}\right]  , \label{V}%
\end{align}
where we have used the boundary conditions to fix most of the integration
constants. The only undetermined integration constant $y_{g}$ will be related
to the chemical potential $\mu$ in the following way. We expand the field $A_{t}\left(  z\right)  $ near the boundary
at $z=0$ to get%
\begin{equation}
A_{t}\left(  0\right)  =\sqrt{\dfrac{-1}{\int_{0}^{z_{H}}y^{3}e^{-3A}%
dy\int_{y_{g}}^{y}\dfrac{x}{e^{A}f}dx}}\left(  -\int_{0}^{z_{H}}\dfrac
{y}{e^{A}f}dy+\dfrac{1}{e^{A\left(  0\right)  }f\left(  0\right)  }%
z^{2}+\cdots\right)  .
\end{equation}
Using the\ AdS/CFT dictionary, we can define the chemical potential in our
system as%
\begin{equation}
\mu=-\sqrt{\dfrac{-1}{\int_{0}^{z_{H}}y^{3}e^{-3A}dy\int_{y_{g}}^{y}\dfrac
{x}{e^{A}f}dx}}\int_{0}^{z_{H}}\dfrac{y}{e^{A}f}dy, \label{mu}%
\end{equation}
in which $y_{g}$ can be solved in term of the chemical potential $\mu$ once
the gauge kinetic function $f\left(  z\right)$ and the warped factor
$A\left(  z\right)$ are given. Put the solution (\ref{phip}-\ref{V}) into
the constraint (\ref{eom-phi}), it is straightforward to verify that the above
solutions are consistent with the constraint.

We note that the solutions (\ref{phip}-\ref{V}) depend on two arbitrary
functions, i.e. the gauge kinetic function $f\left(  z\right)  $ and the
warped factor $A\left(  z\right)  $. Different choices of the functions
$f\left(  z\right)  $ and $A\left(  z\right)  $ will give different physically
allowed backgrounds. Thus we have just found a family of analytic solutions
for the Einstein-Maxwell-dilaton system. We will use the freedom of choosing
functions $f\left(  z\right)  $ and $A\left(  z\right)  $ to satisfy some
extra important physical constraints.

\subsection{Vector Meson Spectrum}

In a theory with linear confinement like QCD, the spectrum of the squared mass
$m_{n}^{2}$\ of mesons is expected to grow as $n$ at zero temperature and zero
density. This is known as the linear Regge trajectories \cite{0507246}.
In the method of AdS/QCD duality, this issue was first addressed in
\cite{0602229} by modifying the dilaton field in the IR region with a $z^{2}$
term, i.e. the soft-wall model. In \cite{0602229}, the $z^{2}$ term was added
by hand to the dilaton field. It means that the fields configuration used in
the soft-wall model is not a solution of the Einstein equations. Dynamically
generating the $z^{2}$ term by consistently solving the Einstein equations has
been considered in several later works \cite{1103.5389,1104.4182} by including
a proper dilaton potential. At finite temperature and density, the temperature
dependent meson spectrum has been studied in
\cite{0703172,0903.2316,0911.2298,1107.2738,1112.5923,1209.4198} with the AdS
thermal gas background replaced by the charged AdS black hole background.

In the previous section, we consistently solved the equations of motion
(\ref{eom-phi}-\ref{eom-V}) for the Einstein-Maxwell-dilaton system. The
analytic solutions depend on two arbitrary functions, the gauge kinetic
function $f\left(  z\right)  $ and the warped factor $A\left(  z\right)  $. In
this section, we will study the meson spectrum in our background and constrain
the functions of $f\left(  z\right)  $ and $A\left(  z\right)  $ by requesting
the vector meson spectrums satisfy the linear Regge trajectories at zero
temperature and zero density.

We consider a 5d probe vector field $V$ whose action has been written down in
(\ref{action-m}),%
\begin{equation}
S_{m}=-\dfrac{1}{16\pi G_{5}}\int d^{5}x\sqrt{-g}{\frac{f\left(  \phi\right)
}{4}F_{V}^{2}}.
\end{equation}
To get the meson spectrum, we study the vector field $V$ in the charged AdS
black hole background which we have obtained in the previous section,%
\begin{align}
ds^{2}  &  =\dfrac{e^{2A\left(  z\right)  }}{z^{2}}\left[  -g(z)dt^{2}%
+\frac{dz^{2}}{g(z)}+d\vec{x}^{2}\right] \label{metric for meson}\\
A_{\mu}  &  =A_{t}\left(  z\right)  dt.
\end{align}
The equation of motion of the vector field $V$ has been derived in
(\ref{eom3}) as%
\begin{equation}
\nabla_{\mu}\left[  f\left(  \phi\right)  F_{V}^{\mu\nu}\right]  ={{0.}}%
\end{equation}
Following \cite{0602229}, we first use the gauge invariance to fix the gauge
$V_{z}=0$, then the equation of motion of the transverse vector field
$V_{\mu}$ $\left(  \partial^{\mu}V_{\mu}=0\right)  $ in the background
(\ref{metric for meson}) reduces to%
\begin{equation}
\dfrac{1}{g}\nabla^{2}V_{i}+V_{i}^{\prime\prime}+\left(  \dfrac{g^{\prime}}%
{g}+\dfrac{f^{\prime}}{f}+A^{\prime}-\dfrac{1}{z}\right)  V_{i}^{\prime}=0,
\label{eqV}%
\end{equation}
where the prime is\ the derivative of $z$. We next perform the Fourier
transformation for the vector field $V_{i}$ as%
\begin{equation}
V_{i}\left(  x,z\right)  =\int\dfrac{d^{4}k}{\left(  2\pi\right)  ^{4}%
}e^{ik\cdot x}v_{i}\left(  z\right)  , \label{V-v}%
\end{equation}
where $k=\left(  \omega,\vec{p}\right)  $ and the functions $v_{i}\left(  z\right)
$\ satisfy the eigen-equations%
\begin{equation}
-v^{\prime\prime}_{i}-\left(  \dfrac{g^{\prime}}{g}+\dfrac{f^{\prime}}{f^{\prime}%
}+A^{\prime}-\dfrac{1}{z}\right)  v^{\prime}_{i}=\left(  \dfrac{\omega^{2}}{g^{2}%
}-\dfrac{p^{2}}{g}\right)  v_{i}. \label{eqv}%
\end{equation}
Redefining the functions $v_{i}\left(  z\right)  $ with%
\begin{equation}
v_{i}=\left(  \dfrac{z}{e^{A}fg}\right)  ^{1/2}\psi_{i}\equiv X\psi_{i},
\end{equation}
brings the equation of motion (\ref{eqv}) into the form of the Schr\"{o}dinger
equation%
\begin{equation}
-\psi^{\prime\prime}_{i}+U\left(  z\right)  \psi_{i}=\left(  \dfrac{\omega^{2}}{g^{2}%
}-\dfrac{p^{2}}{g}\right)  \psi_{i}, \label{Shordinger}%
\end{equation}
where the potential function is%
\begin{equation}
U\left(  z\right)  =\dfrac{2X^{\prime2}}{X^{2}}-\dfrac{X^{\prime\prime}}{X}.
\end{equation}
In the case of zero temperature and zero chemical potential, we expect that
the discrete spectrum of the vector mesons obeys the linear Regge
trajectories. At $\mu=0$, the metric of the zero temperature background
(thermal gas) coincides with the black hole metric in the limit of zero size,
i.e. $z_{H}\rightarrow\infty$, which corresponds to $g\left(  z\right)  =1$.
In the zero size black hole limit, the Schr\"{o}dinger equation reduces to%
\begin{equation}
-\psi^{\prime\prime}_{i}+U\left(  z\right)  \psi_{i}=m^{2}\psi_{i},
\end{equation}
where $-m^{2}=k^{2}=-\omega^{2}+p^{2}$. To produce the discrete mass spectrum
with the linear Regge trajectories, the potential $U\left(  z\right)  $ should
be in certain forms. Following \cite{0602229}, a simple choice is to fix the
gauge kinetic function as $f\left(  z\right)  =e^{\pm cz^{2}-A\left(
z\right)  }$, then the potential becomes%
\begin{equation}
U\left(  z\right)  =-\dfrac{3}{4z^{2}}-c^{2}z^{2}. \label{potential}%
\end{equation}
The Schr\"{o}dinger equations (\ref{Shordinger}) with the above potential
(\ref{potential}) have the discrete eigenvalues
\begin{equation}
m_{n}^{2}=4cn,
\label{mass}%
\end{equation}
which is
linear in the energy level $n$ as we expect for the vector spectrum at zero
temperature and zero density. At finite temperature and finite density,
$g\left(  z\right)  \neq1$, the masses of the vector mesons solved from the
Eq. (\ref{Shordinger}) will depend on the temperature and density. For the
case of small enough temperature and density, Eq. (\ref{Shordinger}) can
be solved perturbatively to get the mass shift from the linear Regge
trajectories \cite{1209.4198,1112.5923,1107.2738,0801.4383}. For large temperature and density, the method of spectral
functions is useful. The study of temperature and density dependent vector
mass spectrum is in progress.

Once we fixed the gauge kinetic function $f=e^{\pm cz^{2}-A\left(  z\right)
}$, the Eq.(\ref{mu}) can be solved to get the integration constant $y_{g}$ in
term of the chemical potential $\mu$ explicitly as%
\begin{equation}
e^{cy_{g}^{2}}=\dfrac{\int_{0}^{z_{H}}y^{3}e^{-3A}e^{cy^{2}}dy}{\int
_{0}^{z_{H}}y^{3}e^{-3A}dy}+\dfrac{\left(  1-e^{cz_{H}^{2}}\right)  ^{2}%
}{2c\mu^{2}\int_{0}^{z_{H}}y^{3}e^{-3A}dy}. \label{yg}%
\end{equation}
Put the integration constant $y_{g}$ back into the solution (\ref{phip}%
-\ref{V}),\ we can finally write down our solution as%
\begin{align}
\phi^{\prime}\left(  z\right)   &  =\sqrt{-6\left(  A^{\prime\prime}%
-A^{\prime2}+\dfrac{2}{z}A^{\prime}\right)  },\label{phip-A}\\
A_{t}\left(  z\right)   &  =\mu\dfrac{e^{cz^{2}}-e^{cz_{H}^{2}}}%
{1-e^{cz_{H}^{2}}},\label{At-A}\\
g\left(  z\right)   &  =1+\dfrac{1}{\int_{0}^{z_{H}}y^{3}e^{-3A}dy}\left[
\dfrac{2c\mu^{2}}{\left(  1-e^{cz_{H}^{2}}\right)  ^{2}}\left\vert
\begin{array}
[c]{cc}%
\int_{0}^{z_{H}}y^{3}e^{-3A}dy & \int_{0}^{z_{H}}y^{3}e^{-3A}e^{cy^{2}}dy\\
\int_{z_{H}}^{z}y^{3}e^{-3A}dy & \int_{z_{H}}^{z}y^{3}e^{-3A}e^{cy^{2}}dy
\end{array}
\right\vert -\int_{0}^{z}y^{3}e^{-3A}dy\right]  ,\\
V\left(  z\right)   &  =-3z^{2}ge^{-2A}\left[  A^{\prime\prime}+3A^{\prime
2}+\left(  \dfrac{3g^{\prime}}{2g}-\dfrac{6}{z}\right)  A^{\prime}-\dfrac
{1}{z}\left(  \dfrac{3g^{\prime}}{2g}-\dfrac{4}{z}\right)  +\dfrac
{g^{\prime\prime}}{6g}\right]  . \label{V-A}%
\end{align}
Now we have fixed all the integration constants by either satisfying the boundary
conditions (\ref{bc-zH}) and (\ref{bc-0}) or relating to the chemical potential $\mu$. The final solution (\ref{phip-A}%
-\ref{V-A}) depends only on the warped factor $A\left(  z\right)  $. The
choice of $A\left(  z\right)  $ is arbitrary provided it satisfies the
boundary condition (\ref{bc-0}). In the next sections, we will make a simple
choice of $A\left(  z\right)  $ and use it to study the phase structure in its
holographic QCD model.%

\setcounter{equation}{0}
\renewcommand{\theequation}{\arabic{section}.\arabic{equation}}%

\section{Phase Structure}

We will study the phase structure for the black hole background which we
obtained in the last section (\ref{phip-A}-\ref{V-A}). The phase transitions
in the black hole background correspond to the phase transitions in its
holographic QCD theory by AdS/QCD duality.

\subsection{Fixing the Warped Factor }

To be concrete, we fix the warped factor $A\left(  z\right)  $ in our solution
in a simple form as%
\begin{equation}
A\left(  z\right)  =-\dfrac{c}{3}z^{2}-bz^{4}, \label{A}%
\end{equation}
where the parameters $b$ and $c$ will be determined by later. It is easy to
show that this choice of $A\left(  z\right)  $ satisfies the boundary
condition (\ref{bc-0}) by Eq. (\ref{phip-A}). There are many more complicated
choices for the function of $A\left(  z\right)  $, but we will show that our
simple choice already educes abundant phenomena in QCD.

Once we choose the function of $A\left(  z\right)  $ up to the parameters $b$
and $c$, our solution (\ref{phip-A}-\ref{V-A}) is completely fixed. Expanding
the dilaton field and the dilaton potential near the boundary $z=0$, respectively,%

\begin{align}
\phi\left(  z\right)   &  =2\sqrt{3c}z+\dfrac{2\left(  c^{2}+30b\right)
}{9\sqrt{3c}}z^{3}+\cdots,\\
V  &  =-12-18cz^{2}+\cdots,
\end{align}
we can write the dilaton potential in terms of the dilaton field $\phi$ as the
expected form from the AdS/CFT dictionary,
\begin{equation}
V=-12+\dfrac{\Delta\left(  \Delta-4\right)  }{2}\phi^{2}+\cdots,\Delta=3.
\end{equation}
The conformal dimension $\Delta=3$ satisfies the BF bound $2<\Delta<4$ implying that our gravitational background is stable. Furthermore, the dilaton satisfying the BF bound corresponds to a
local, gauge invariant operator in 4d QCD possibly. One should
note that the dilaton in our work is different from that in the
references \cite{1201.0820,1103.5389,1209.4512}, where the dictionary between the dilaton and gauge invariant operator in field theory are not clear. In our case, we combine the effects of quarks and gluon which absorbed by dilaton
potential.

We fix the parameter $c$ by fitting our mass formula\footnote{We shift our mass formula (\ref{mass}) by 1 to make $n=1$ correspond to the lowest quarkonium states, $J/\psi$.} $m_{n}^{2}=4c\left(
n+1\right)$ to the lowest two quarkonium states\footnote{According to the
analysis of the phase structure of our background that we will see later in
this section, we would like to interpret our holographic QCD model to describe
the heavy quarks system.},
\begin{equation}
m_{J/\psi}=3.096GeV \text{,\ \ } m_{\psi^{\prime}}=3.685GeV,
\end{equation}
For $c\simeq 1.16 GeV^2$, we have%
\begin{equation}
m_{1}=3.046GeV \text{,\ \ } m_{2}=3.731GeV,
\end{equation}
which are consistent with the experimental data within $1\%$. The
parameter $b$ will be determined later by comparing the the lattice QCD result of the phase transition temperature at zero chemical potential.

\subsection{Black Hole Thermodynamics}

Using the black hole metric we obtained%
\begin{equation}
ds^{2}=\dfrac{e^{2A\left(  z\right)  }}{z^{2}}\left[  -g(z)dt^{2}+\frac
{dz^{2}}{g(z)}+d\vec{x}^{2}\right]  ,
\end{equation}
it is easy to calculate the Hawking-Bekenstein entropy%
\begin{equation}
s=\dfrac{area\left(  z_{H}\right)  }{4}=\dfrac{e^{3A\left(  z_{H}\right)  }%
}{4z_{H}^{3}}, \label{entropy}%
\end{equation}
and the Hawking temperature%
\begin{equation}
T=\dfrac{\kappa}{2\pi}=\dfrac{z_{H}^{3}e^{-3A\left(  z_{H}\right)  }}{4\pi
\int_{0}^{z_{H}}y^{3}e^{-3A}dy}\left[  1-\dfrac{2c\mu^{2}\left(  e^{cz_{H}%
^{2}}\int_{0}^{z_{H}}y^{3}e^{-3A}dy-\int_{0}^{z_{H}}y^{3}e^{-3A}e^{cy^{2}%
}dy\right)  }{\left(  1-e^{cz_{H}^{2}}\right)  ^{2}}\right]  .
\end{equation}

\begin{figure}
[h]
\begin{center}
\includegraphics[
height=2in,
width=3in
]%
{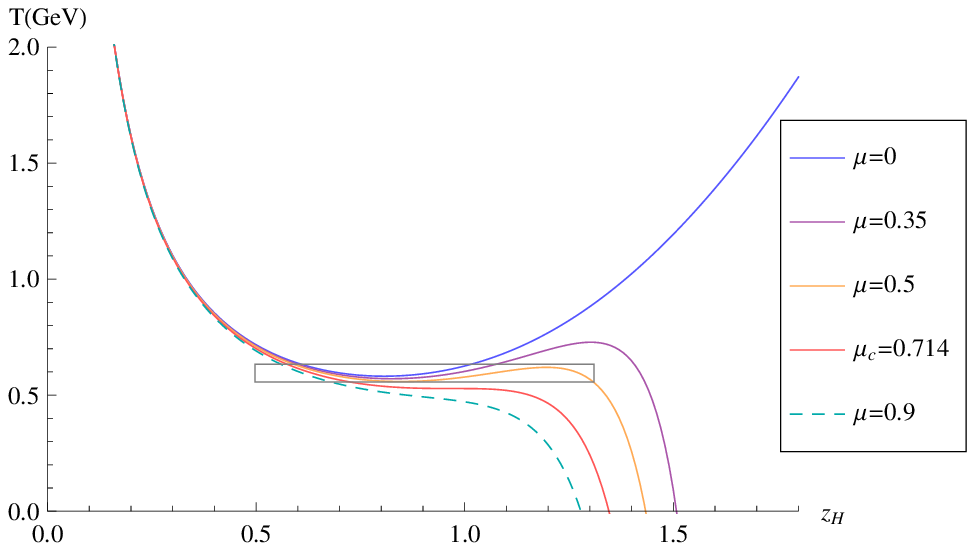}%
\hspace*{1cm}
\includegraphics[
height=2in,
width=2.9in
]%
{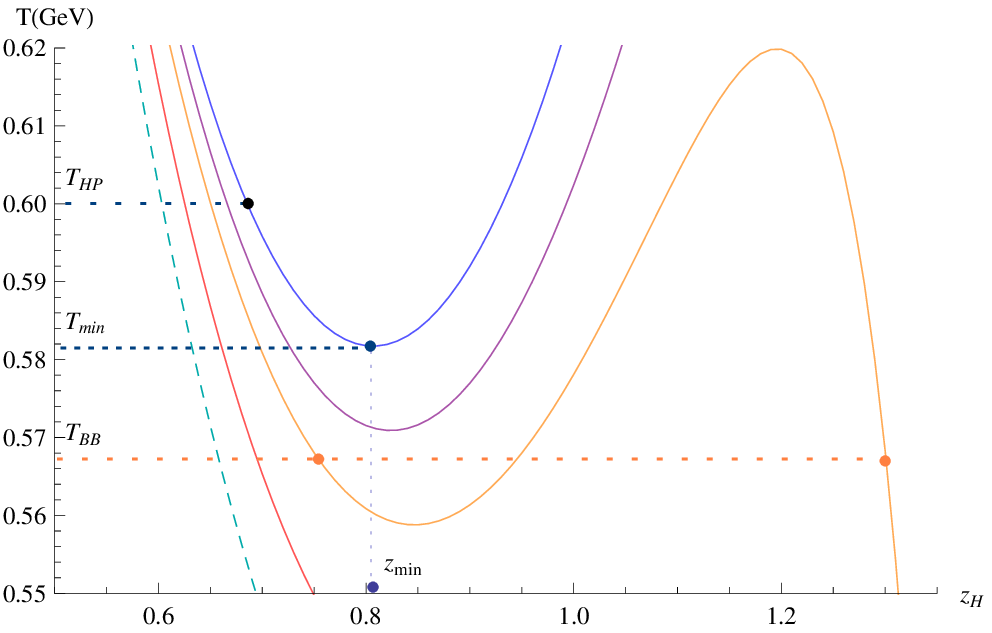}%
\vskip -0.05cm \hskip 0.15 cm
\textbf{( a ) } \hskip 7.5 cm \textbf{( b )}
\caption{The temperature v.s. horizon at the different chemical potentials
$\mu=0,0.35,0.5,0.714,0.9GeV$ are plotted. We enlarge the rectangle region in (a) into (b) to see the detailed structure. At $\mu=0$, there is a global minimum $T_{\min}$; for $0<\mu<\mu_{c}\simeq0.714GeV$, the minimum becomes local and eventually disappears for $\mu\geq\mu_{c}$.}%
\label{temperature}%
\end{center}
\end{figure}

We plot the temperature $T$ v.s. horizon $z_{H}$ at different chemical
potentials in FIG. \ref{temperature}. At $\mu=0$, the temperature has a
global minimum $T_{\min}$ at $z_{H}=z_{\min}$. For $z_{H}>z_{\min}$, the black
hole solutions are thermodynamically unstable. Below the temperature $T_{\min
}$, there is no black hole solution and we expect a Hawking-Page phase
transition happens at a temperature $T_{HP}\gtrsim T_{\min}$ where the black
hole background transits to a thermal gas background. For $0<\mu<\mu_{c}%
\simeq 0.714 GeV$, the temperature has a local minimum/maximum temperature $T_{\mu\min
}/T_{\mu\max}$\ at $r_{H}=r_{\min}/r_{\max}$ and decreases to zero at a finite
size of horizon. The black holes between $r_{\min}$ and $r_{\max}$ are
thermodynamically unstable. We expect a similar Hawking-Page phase transition
happens at a temperature $T_{HP}\gtrsim T_{\mu\min}$. In addition, since the
thermodynamically stable black hole solutions exist even when the temperature
below $T_{\mu\min}$, we also expect a black hole to black hole phase
transition happens at a temperature $T_{BB}\in\left[  T_{\mu\min},T_{\mu\max
}\right]  $, where the large black hole transits to a small black hole.
The\ values of $T_{HP}$ and $T_{BB}$ will determine the true vacuum state,
thermal gas or small black hole, in which the system will stay eventually.
Finally, for $\mu>\mu_{c}$, the temperature monotonously decreases to zero and
there is no black hole to black hole phase transition anymore\footnote{There
could still be a Hawking-Page phase transition at some temperature for the
case of $\mu>\mu_{c}$, but we will show later that the black hole solution is
always thermodynamically favored in the case.}, which implies that there is a
second order phase transition happens at $\mu=\mu_{c}$, i.e. the critical point.%

To determine the phase transition temperatures $T_{HP}$ and $T_{BB}$, we
compute the free energy from the first law of thermodynamics in grand
canonical ensemble,%
\begin{equation}
F=\epsilon-Ts-\mu\rho.
\end{equation}
Changes in the free energy of a system with constant volume are given by%
\begin{equation}
dF=-sdT-\rho d\mu.
\end{equation}
At fixed values of the chemical potential $\mu$, the free energy can be
evaluated by the integral \cite{0812.0792}%
\begin{equation}
F=-\int sdT. \label{int F}%
\end{equation}
We can fix the integration constant in the above integral (\ref{int F}) by
considering the zero chemical potential case. At $\mu=0$, the metric of the zero temperature background (thermal gas) coincides with the black hole metric
in the limit of zero size, i.e. $z_{H}\rightarrow\infty$, where we expect that
the free energy of the black hole background also coincides with the free
energy of the thermal gas background which we can choose to be zero. Thus we
require $F\left(  z_{H}\rightarrow\infty\right)  =0$ and obtain that%
\begin{equation}
F=\int_{z_{H}}^{\infty}s\dfrac{dT}{dz_{H}}dz_{H}. \label{F}%
\end{equation}

\begin{figure}
[h]
\begin{center}
\includegraphics[
height=2.2in,
width=3in
]%
{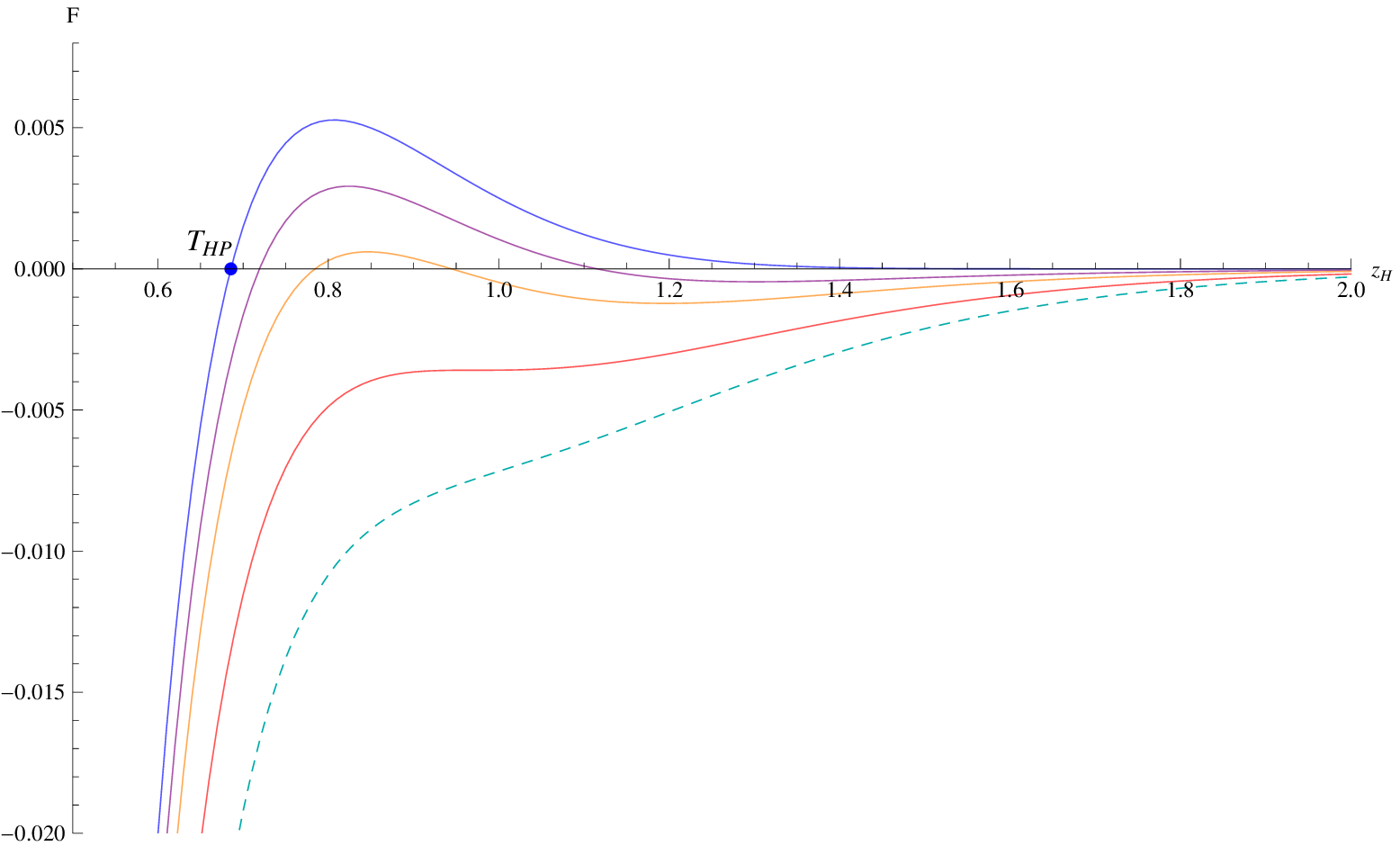}%
\hspace*{0.6cm}
\includegraphics[
height=2.2in,
width=3in
]%
{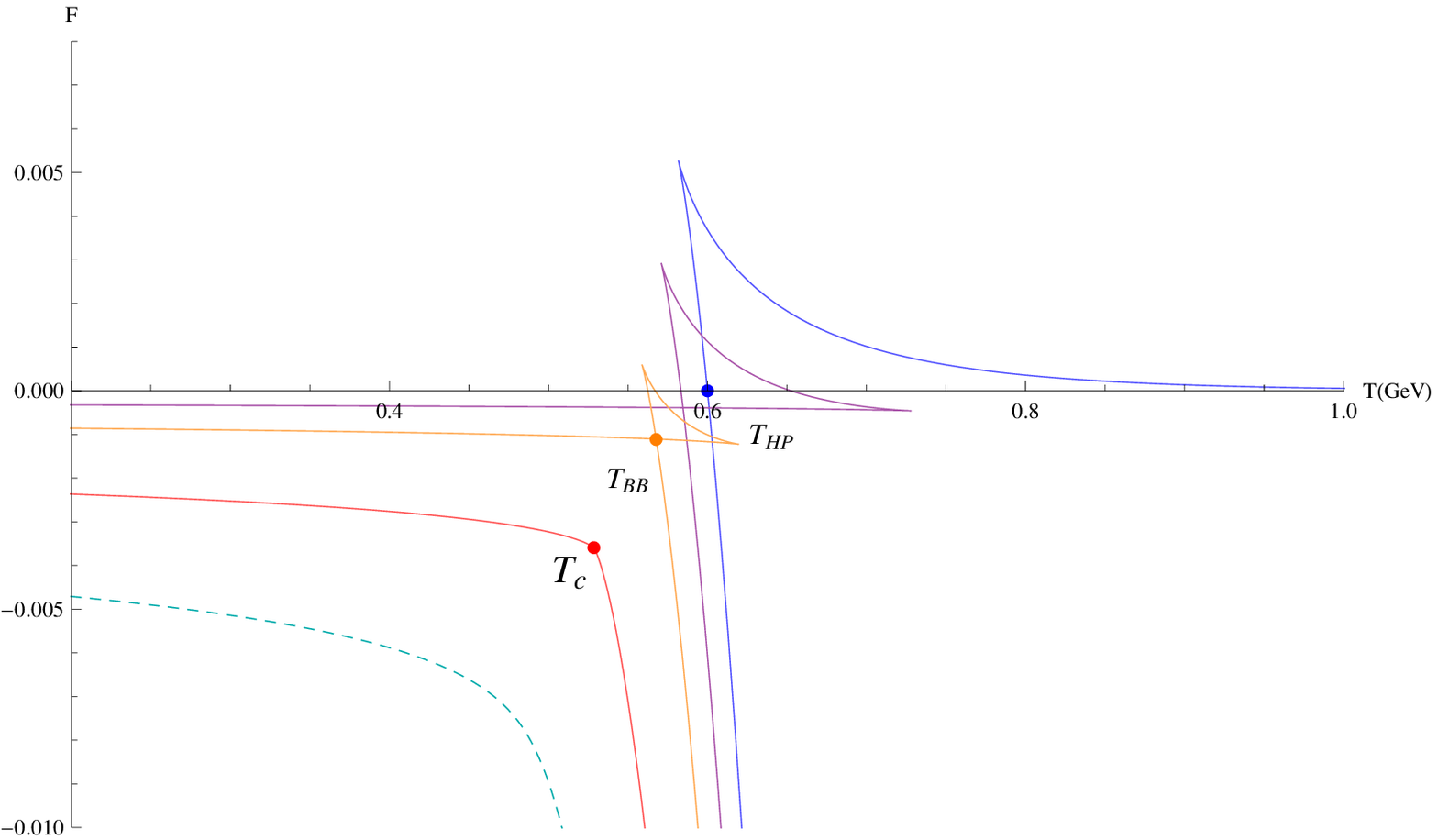}%
\vskip -0.05cm \hskip 0.15 cm
\textbf{( a ) } \hskip 6.5 cm \textbf{( b )}
\caption{(a) The free energy v.s. horizon at the different chemical potentials
$\mu=0,0.35,0.5,0.71,0.9$. At $\mu=0$, we normalized that the free energy
vanishes when $z_{H}\rightarrow\infty$. For $\mu<\mu_{c}\simeq0.71$, the free
energy has a maximum; for $\mu\geq\mu_{c}$, the free energy becomes
monotonous. (b) The free energy v.s. temperature at the different chemical potentials. At $\mu=0$, we normalized that the free energy
vanishes when $z_{H}\rightarrow\infty$. For $\mu<\mu_{c}$, the free
energy has a maximum; for $\mu\geq\mu_{c}$, the free energy becomes
monotonous.}%
\label{free energy}%
\end{center}
\end{figure}

With the choice of $A\left(  z\right)  $ in (\ref{A}), the integral in
(\ref{F}) can be performed to get the free energy of the black hole at
fixed chemical potentials. We plot the free energy v.s. horizon at different
chemical potentials in FIG. \ref{free energy}. We have normalized the free energy of
the black holes by requiring it to vanish (or equal to the free energy of the thermal gas) at $z_{H}\rightarrow\infty$. Therefore, the black hole background is favored for the negative value of the free energy and the thermal gas background is favored for the positive value of the free energy. At $\mu=0$,
the free energy starts from a large negative value at small $z_{H}$ to a positive maximum value and then decreases to zero at $z_{H}\rightarrow\infty$.
The free energy intersecting the $x$-axis implies that there exists a Hawking-Page phase transition from the black hole to
the thermal gas background at $T_{HP}$. For $0<\mu<\mu_{c}$, besides a maximum
value, the free energy has a local minimum value, which implies a black hole
to black hole phase transition. For $\mu\geq\mu_{c}$, the free energy becomes
monotonous and no phase transition exists.%

The phase structure is more transparent in the plot of free energy v.s.
temperature in FIG. \ref{free energy}. At $\mu=0$, the free energy increase from a
negative value with a large temperature to zero at $T=T_{HP}$ where the black
hole transits to the thermal gas background which is thermodynamically stable
for $T<T_{HP}$. At finite $\mu$, the free energy behaves as the expected
swallow-tailed shape. For $0<\mu<\mu_{c}$, the curve of free energy intersects
with itself at $T=T_{BB}$ where the large black hole transits to the small
black hole background. We found that the free energy of the black hole is
always less than that of the thermal gas, i.e. $F_{\text{black hole}%
}<F_{\text{thermal gas}}=0$. Therefore the thermodynamic system will always
favor the small black hole background other than the thermal gas background
when $T<T_{BB}$. When we increase the chemical potential $\mu$ from zero to
$\mu_{c}$, the loop of the swallow-tailed shape shrinks to disappear at
$\mu=\mu_{c}$, where the background undergoes a second phase transition. For
$\mu>\mu_{c}$, the curve of the free energy increases smoothly from higher
temperature to lower temperature.

\begin{figure}
[h]
\begin{center}
\includegraphics[
height=2in,
width=3in
]%
{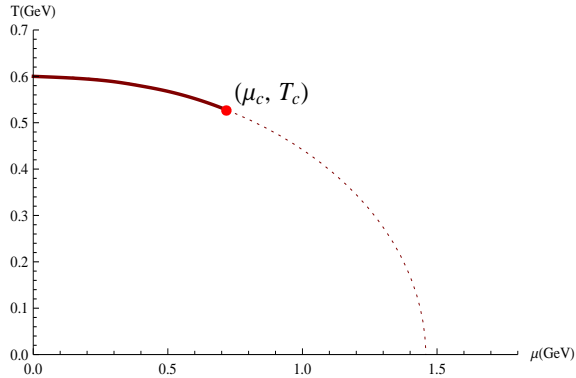}%
\caption{The phase diagram in $T$ and $\mu$ plane. At small $\mu$, the system
undergoes a first order phase transition at finite $T$. The first order phase
transition stops at the critical point $\left(\mu_{c},T_{c}\right) \simeq \left(0.714GeV,0.528GeV \right)$, where
the phase transition becomes second order. For $\mu>\mu_{c}$, the system
weaken to a sharp but smooth crossover.}%
\label{phase diagram}%
\end{center}
\end{figure}

We plot the phase diagram of our holographic QCD model in FIG. \ref{phase diagram}. At $\mu=0$, the system undergoes a black hole to
thermal gas first order phase transition at $T=T_{HP}$. For $0<\mu<\mu_{c}$,
the system undergoes a large black hole to small black hole first order phase
transition at $T_{BB}$. The phase transition temperature $T_{BB}$ approaches
to $T_{HP}$ at $\mu\rightarrow0$ that makes the phase diagram continuous at
$\mu=0$. By comparing the phase transition temperature at $\mu=0$ to the lattice QCD simulation of $T_{HP}\simeq 0.6GeV$ in \cite{1111.4953}, we fix the parameter $b\simeq 0.273GeV^4$. The first order phase transition stops at the critical point $\left(\mu_{c},T_{c} \right)$, where the phase transition becomes second order. For $\mu>\mu_{c}$, the system has a sharp but smooth crossover.

The phase diagram of our holographic QCD model in FIG. \ref{phase diagram} is
not expected from the common picture of QCD phase diagram, in which
the\ transition is crossover for the small chemical potential $\mu$ and
sharpens to the first order phase transition beyond a critical $\mu_{c}$.
Then question is how to interpret our result? To explain our result, we need
to look at the phase structure of QCD more carefully. Lattice QCD provides many
useful information about the phase diagram of QCD at least at $\mu=0$. For
finite density, lattice QCD suffers the well-known sign problem. However,
several perturbative methods such as reweighting, complex chemical potential
and expansions in $\mu/T$ etc. have been developed in lattice QCD to compute
the physical quantities at finite density. Despite that these methods only
work for small chemical potential $\mu$, they can help us to understand the whole
picture of QCD phase diagram.

\begin{figure}
[h]
\begin{center}
\includegraphics[
height=2.7in,
width=3in
]%
{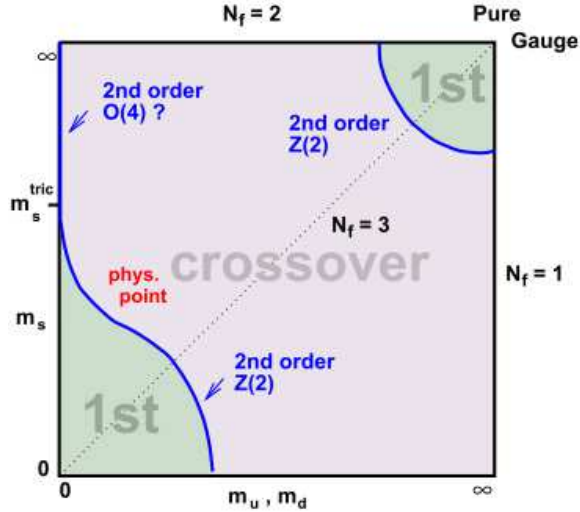}%
\caption{Schematic phase transition behavior of three flavor QCD for
different choices of quark masses at zero density \cite{1009.4089}.}%
\label{mu=0}%
\end{center}
\end{figure}

FIG. \ref{mu=0} shows the schematic phase transition behavior of three
flavor QCD for different choices of quark masses at $\mu=0$ \cite{1009.4089}.
We can see that, the order of the phase transition at $\mu=0$ depends on the
quark masses\footnote{Most of the lattice results prefer that the physical
mass point locates in the crossover region, but this is not completely
confirmed yet \cite{0607017}.}. There are two limit regions where the phase
transition is first order at $\mu=0$. One of them is near the chiral limit,
i.e. $m_{u}=m_{d}=m_{s}=0$. The other is near the decoupling limit, i.e.
$m_{u}=m_{d}=m_{s}\rightarrow\infty$. Thus there are two possibilities to
interpret the phase diagram of our holographic QCD model.

\begin{figure}
[h]
\begin{center}
\includegraphics[
height=1.6in,
width=6in
]%
{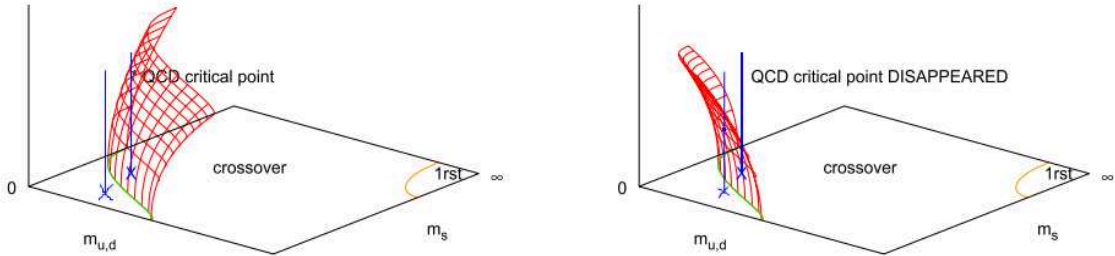}%
\caption{The chiral critical surfaces in the case of positive (left) and
negative (right) curvature from lattice QCD \cite{1009.4089}.}%
\label{curvatures}%
\end{center}
\end{figure}

The first possibility is to interpret our result as for the light quarks (near
the point of chiral limit) and the phase transition is the chiral symmetry
breaking phase transition. For a not very large chemical potential $\mu$,
lattice calculation shows that the chiral critical line separating the
crossover from the first order phase transition regions expands as increasing
the chemical potential $\mu$ to form a chiral critical surface. The sign of
the curvature of the critical surface is crucial to determine the phase
structure at finite $\mu$. In FIG. \ref{curvatures}, the chiral critical
surfaces in the case of positive and negative curvature are showed. In the
interpretation of the light quarks, the mass point locates near the origin.
The first order transition will be preserved for any finite $\mu$ if the
curvature of the chiral critical surface is positive, while it will becomes a
crossover at a finite chemical potential $\mu_{c}$ if the curvature of the
chiral critical surface is negative. Recent lattice calculations shows that
the curvature of the chiral critical surface is more like negative
\cite{0607017,1009.4089} that is consistent with the phase diagram of our
holographic QCD model.%

\begin{figure}
[h]
\begin{center}
\includegraphics[
height=2.3in,
width=3.5in
]%
{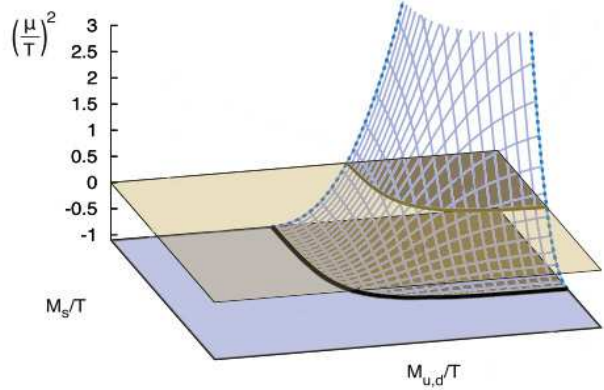}%
\caption{The deconfinement critical surface from lattice QCD \cite{1111.4953}.}%
\label{deconfinement curvature}%
\end{center}
\end{figure}

The other possibility is to interpret our result as for the heavy quarks (near
the point of decoupling limit) and the phase transition is the deconfinement
phase transition. Similar to the chiral critical line, lattice calculation
shows that the deconfinement critical line separating the crossover from the
first order phase transition regions expands as increasing the chemical
potential $\mu$ to form a deconfinement critical surface. FIG.
\ref{deconfinement curvature} shows the recent result of the deconfinement
critical surface from lattice QCD \cite{1111.4953}. In our interpretation of
heavy quarks, the mass point locates near the infinite mass point. The
first order phase transition at $\mu=0$ will becomes a crossover at a finite
chemical potential $\mu_{c}$ and it is consistent with the phase diagram of
our holographic QCD model. The phase transition for heavy quarks has been
extensively studied recently by using lattice QCD method
\cite{1009.4089,1011.4747,1106.0974,1111.4953,1202.6113,1207.3005,1210.7994}.

Among the two possible interpretations of our holographic QCD model, \ we will
focus on the heavy quark interpretation in the rest of our paper. The reason
is that, for the light quarks, the lattice QCD simulation has almost confirmed
that the physical point of light quarks mass locates in the crossover region
at zero chemical potential in FIG. \ref{mu=0}. While for the heavy quarks, the
lattice QCD does not give any constraint yet and the physical point of the
heavy quarks mass has great possibility to locate in the first order region at
zero chemical potential. Furthermore, if we re-examine the gravity side more
carefully, we realize that we did not take into account the backreaction from
the matter fields when we solved our gravitational background. We only
consider the baryon number chemical potential, but not the dynamical quarks.
This means that we are considering the quenched limit of the heavy quarks.

To interpret our result as for the heavy quarks with the deconfinement phase
transition, there is still a problem in the gravity side. It is widely
believed that the deconfinement phase transition in the field theory side is
dual to the Hawking-Page phase transition in the gravity side. Hawking-Page
phase transition is the transition between black hole and thermal gas
backgrounds. However, in our gravity background, the phase transition is
between a large black hole and a small black hole backgrounds for non-zero
chemical potential. Thus it sounds inconsistent to consider that the black
hole to black hole phase transition in the gravity side is dual to the
deconfinement phase transition in QCD. Our resolution for this problem is
that, although it is thermodynamically stable, the small black hole is
dynamical unstable. On the other hand, the
gauge group in realistic QCD is $SU\left(  N\right)  \sim SU\left(
3\right)  $. Thus we have to consider the finite $N$ effect in the gravity
side. If we still require that the gravity limit held, this is merely to
consider the finite string coupling constant by including the loop correction,
i.e. the quantum effect. It is well-known that small black holes are quantum
mechanically unstable and will quickly evaporate away by quantum radiation.
Therefore, right after the phase transition of a large black hole to a small
black hole, the small black hole will continue to evaporate away quickly to a
thermal gas background. In this sense, the black hole to black hole phase
transition can be interpreted as the deconfinement phase transition in the
dual QCD theory.

In our heavy quarks interpretation, an important question is where the
critical point is, i.e. the point where the first order phase transition cease
to become a crossover in the phase diagram FIG.
\ref{phase diagram}. Locating the critical point is a crucial job to
understand the phase structure of QCD. As we have seen in FIG.
\ref{free energy}, the critical point is the point where the self-intersection
disappears. We thus obtain the critical point as $\left(\mu_{c},T_{c}\right) \simeq \left(0.714GeV,0.528GeV \right)  $. To justify the critical point we
got, we expand the the function $A_{t}\left(  z\right)  $ in Eq.(\ref{At-A})
near the boundary $z\rightarrow0$ as%
\begin{equation}
A_{t}\left(  z\right)  =\mu\dfrac{e^{cz^{2}}-e^{cz_{H}^{2}}}{1-e^{cz_{H}^{2}}%
}=\mu+\dfrac{2c\mu}{1-e^{cz_{H}^{2}}}z^{2}+\cdots,
\end{equation}
from which the baryon $\rho$\ density\ can be read off as%
\begin{equation}
\rho=-\dfrac{2c\mu}{1-e^{cz_{H}^{2}}}.
\end{equation}

\begin{figure}
[h]
\begin{center}
\includegraphics[
height=2in,
width=3in
]%
{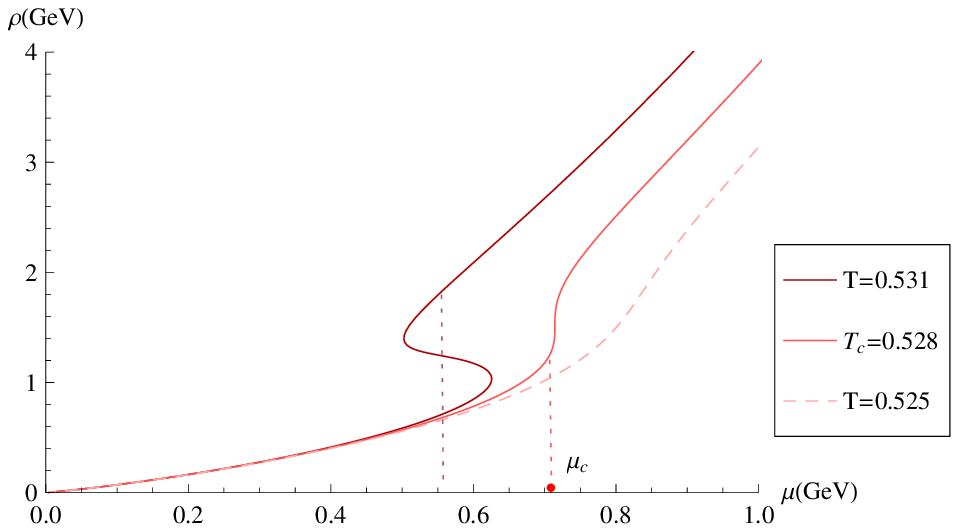}%
\hspace*{0.7cm}
\includegraphics[
height=2in,
width=3in
]%
{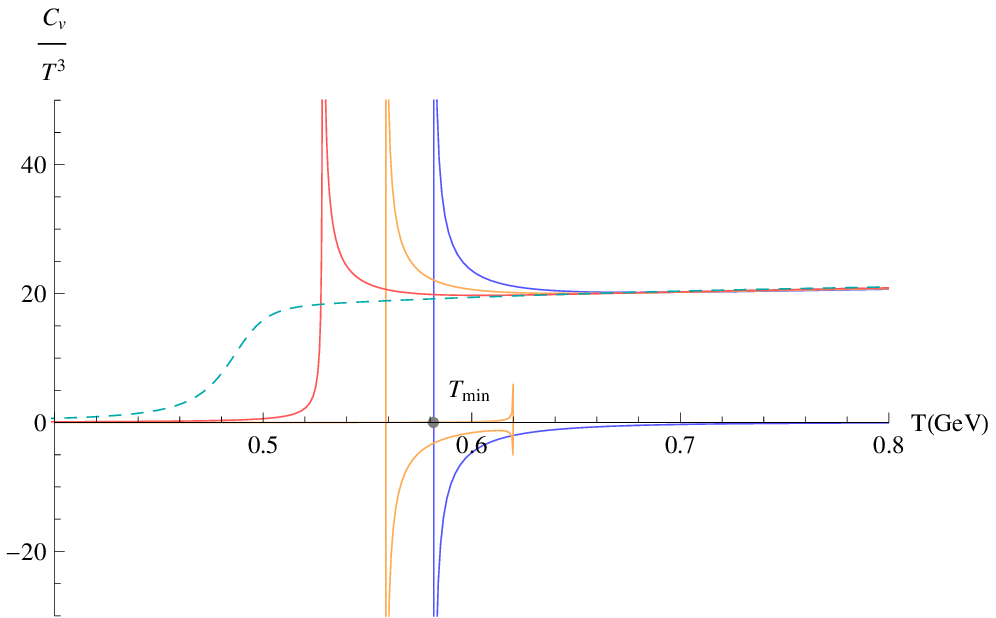}%
\vskip -0.05cm \hskip 0.15 cm
\textbf{( a ) } \hskip 7.5 cm \textbf{( b )}
\caption{(a) The baryon density $\rho$ v.s. the chemical potential $\mu$. The curve is single-valued for
$T>T_{c}$, while it is multi-valued for $T<T_{c}$. At $T=T_{c}$, the
slope becomes infinite at $\mu=\mu_{c}$. (b) The specific heat $C_{V}$ v.s. Temperature. The negative value of $C_{V}$ corresponds to the thermodynamically instability. For $\mu<\mu_{c}$, $C_{V}$ is negative for a range of $T$. For $\mu>\mu_{c}$, the specific heat is always positive implying no phase transition.}%
\label{rho-mu and C}%
\end{center}
\end{figure}

We plot the baryon density $\rho$ v.s. the chemical potential $\mu$ in (a) of FIG.
\ref{rho-mu and C}. For $T<T_{c}$, the baryon density $\rho$\ is single-valued which
indicates that there is no phase transition. While for $T>T_{c},$ $\rho$\ is
multi-valued which indicates that there is a phase transition at certain value
of the chemical potential $\mu$. At the critical temperature $T=T_{c}$, the slope of the $\rho-\mu$ curve becomes infinite at the
critical chemical potential $\mu_{c}$. The behavior of the baryon
number $\rho\left(  \mu\right)  $ near the critical temperature $T_{c}$\ is
consistent with the result we obtained from the free energy.

The susceptibility is defined as%
\begin{equation}
\chi=\left(  \dfrac{\partial\rho}{\partial\mu}\right)  _{T},
\end{equation}
which is just the slope in (a) of FIG. \ref{rho-mu and C}. For $T<T_{c}$, the
susceptibility is always positive, $\chi>0$ implies that the black hole with
any chemical potential value is thermodynamically stable. On the other hand,
for $T<T_{c}$, $\chi$ could be negative for a range of $\mu$ where the black
hole is thermodynamically unstable. At $T=T_{c}$, $\chi\rightarrow\infty$ at
$\mu=\mu_{c}$ which indicates that a phase transition happens around there.

The similar behavior of the susceptibility $\chi$ can be seen from the plot of
the specific heat $C_{V}$ v.s. the temperature $T$ in (b) of FIG. \ref{rho-mu and C}, where the
specific heat $C_{V}$ is defined as%
\begin{equation}
C_{V}=T\left(  \dfrac{\partial s}{\partial T}\right)  _{\mu}.
\end{equation}
We note that, in the $C_{V}-T$ diagram, the negative value of the specific heat
corresponds to the thermodynamically instability. For $\mu>\mu_{c}$, the
specific heat is always positive, $C_{V}>0$ implies that the black hole with any
temperature is thermodynamically stable. While for $0<\mu<\mu_{c}$, $C_{V}$ could
be negative for a range of $T$ where the black hole is thermodynamically
unstable. At $\mu=0$, there is a minimum temperature $T_{\min}$ for the black
hole solutions where the specific heat diverges. The Hawking-Page like phase
transition happens at a temperature slightly above $T_{\min}$ at $T_{HP}$.

\subsection{Equations of State}

The speed of sound is defined as%
\begin{equation}
c_{s}^{2}=\dfrac{\partial\ln T}{\partial\ln s}.
\end{equation}

\begin{figure}
[h]
\begin{center}
\includegraphics[
height=2.3in,
width=3in
]%
{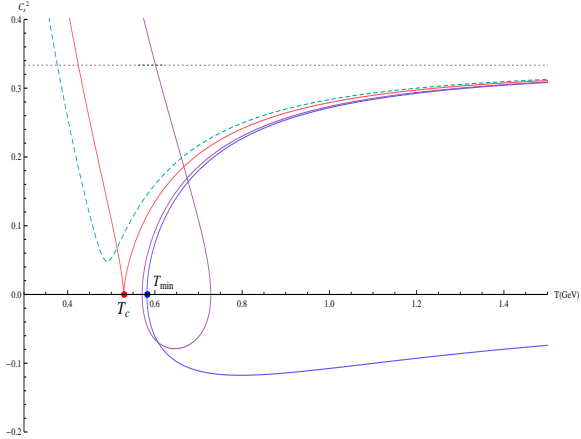}%
\caption{The squared of speed of sound $c_{s}^{2}$ v.s. temperature. For $\mu<\mu_{c}$, $c_{s}^{2}$ is negative for a range of $T$ implying dynamical instability and there is a phase transition. For $\mu>\mu_{c}$, $c_{s}^{2}$ becomes smooth and always positive. At $\mu=\mu_{c}$, $c_{s}^{2}$ touches the $x$-axis at $T=T_c$ where the phase transition transfers to a crossover.}%
\label{cs-T}%
\end{center}
\end{figure}

FIG. \ref{cs-T} plots the squared of speed of sound $c_{s}^{2}$ v.s. the
temperature $T$. For $0<\mu<\mu_{c}$, the speed of sound is imaginary for a
range of temperature, indicating a Gregory-Laflamme instability
\cite{9301052,9404071}. This is related to the general version of Gubser-Mitra
conjecture \cite{0009126,0011127,0104071}, i.e. the dynamical stability of a
horizon is equivalent to the thermodynamic stability. In our system, the
negative specific heat implies thermodynamically unstable. While the imaginary
speed of sound implies the amplitude of the fixed momentum sound wave would
increase exponentially with time, reflecting the dynamical instability. Roughly
speaking, $C_{V}<0$ is equivalent to $c_{s}^{2}<0$ in our system. For $\mu>\mu
_{c}$, the speed of sound behaves as a sharp but smooth crossover. At the
critical point $\mu=\mu_{c}$, a second order phase transition happens where
$c_{s}^{2}$ goes to $0$ at the critical temperature $T_{c}$ but never becomes
negative. In all the case, $c_{s}^{2}$ approaches the comformal limit $1/3$ at
very high temperature as expected.

\begin{figure}
[h]
\begin{center}
\includegraphics[
height=1.8in,
width=2.7in
]%
{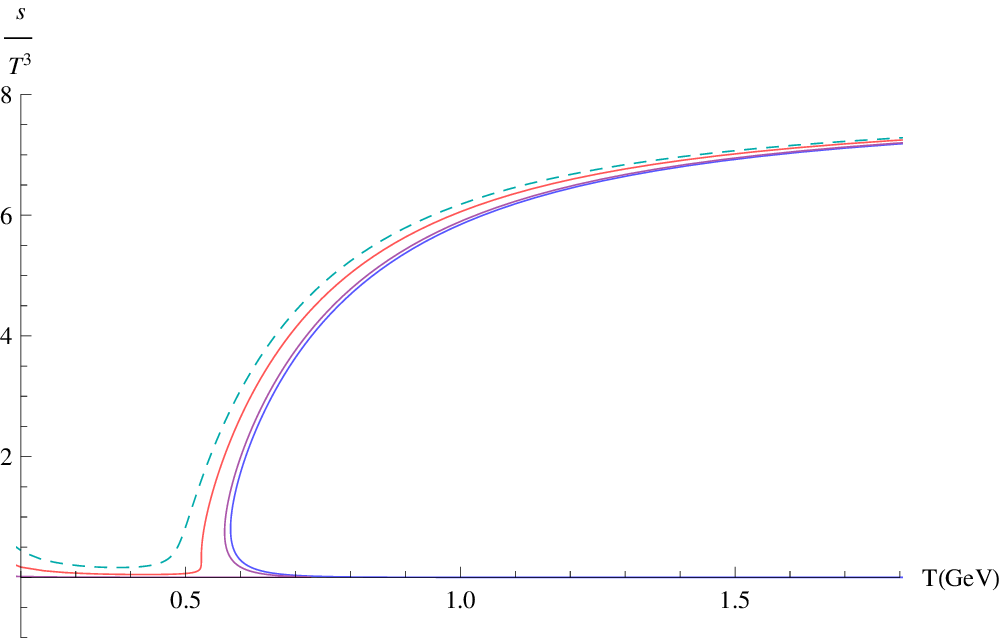}%
\hspace*{0.6cm}
\includegraphics[
height=1.8in,
width=2.7in
]%
{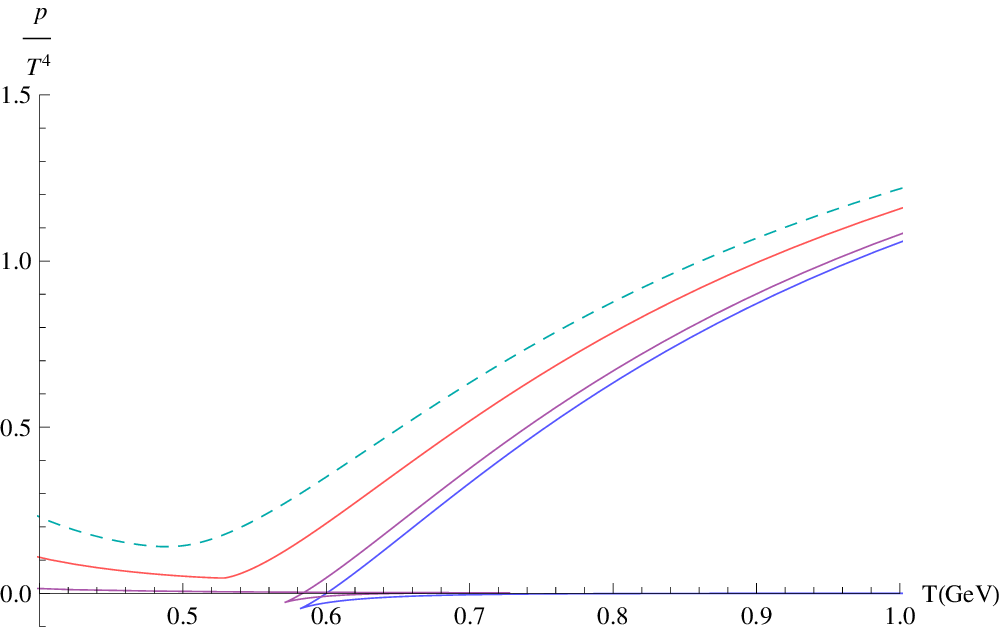}%
\vskip -0.05cm \hskip 0.15 cm
\textbf{( a ) } \hskip 6.5 cm \textbf{( b )}
\end{center}
\begin{center}
\includegraphics[
height=1.8in,
width=2.7in
]%
{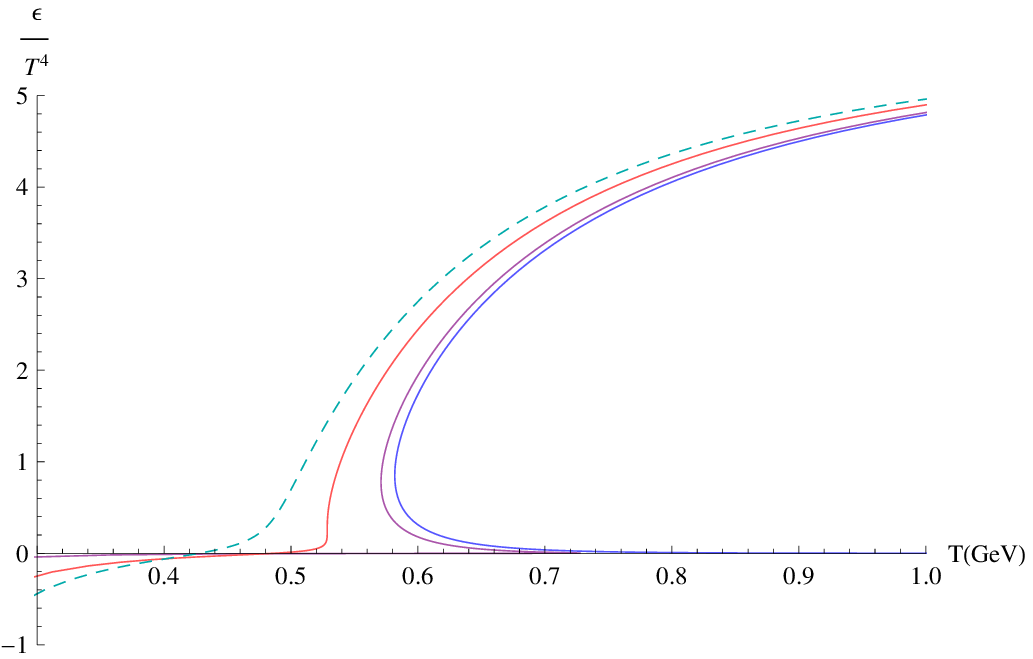}%
\hspace*{0.6cm}
\includegraphics[
height=1.8in,
width=2.7in
]%
{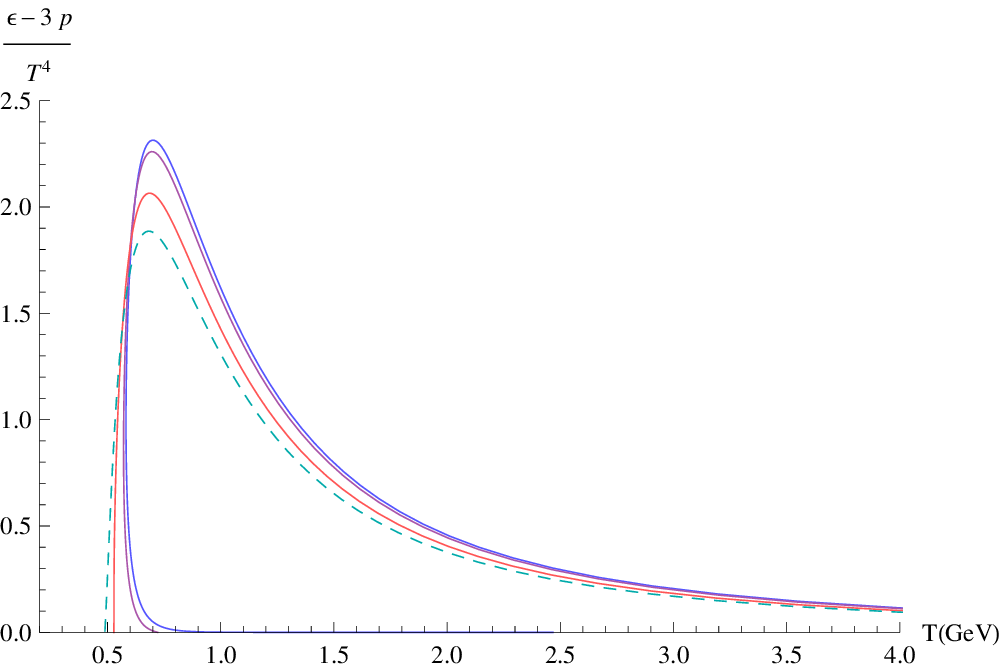}%
\vskip -0.05cm \hskip 0.15 cm
\textbf{( c) } \hskip 6.5 cm \textbf{( d )}
\caption{(a) The entropy $s$ v.s. temperature. (b) The pressure $p$ v.s. temperature. (c)The energy $\epsilon$ v.s. temperature. (d) The trace anomaly $\epsilon-3p$ v.s. temperature.}%
\label{EoS}%
\end{center}
\end{figure}

We plot other equations of state in FIG. \ref{EoS}. The entropy of our black hole solution has been calculated in (\ref{entropy}) and is plotted in (a) of FIG. \ref{EoS}. At $\mu=0$, there is a minimum temperature $T_{\min}$ for the black hole solutions. The
black hole solutions with very low entropy and high temperature always have
negative specific heat and are thermodynamically unstable and the black hole
will transit to the thermal gas through a Hawking-Page phase transition. For
$0<\mu<\mu_{c}$, the entropy is multi-valued for a region of temperature which
indicates a phase transition between high entropy and low entropy black holes.
For $\mu\geq\mu_{c}$, the entropy is single-valued and there is no phase
transition. The similar phase behaviors have been discussed in
\cite{0804.0434} for a holographic QCD model with different values of
parameters tuned by hand.

The pressure $p$\ and the energy $\epsilon$ can be calculated from the free
energy as%
\begin{equation}
p=-F\text{, \ \ }\epsilon=F+sT.
\end{equation}
(b) and (c) of FIG. \ref{EoS} plot the pressure $p$\ and the energy $\epsilon$
v.s. temperature $T$. For some particular $\mu$ and $T$, the pressure becomes
negative which seems not make sense for a real physical system. However, we
see that the pressures for all the thermodynamically favorite backgrounds are
always positive. Both the pressure and the energy increase with the chemical
potential, that pushes the phase transition temperature $T_{BB}$ to the
smaller values for growing $\mu$. Our results are consistent to the recent
lattice results with finite chemical potential \cite{1204.6710}.

We finally plot the trace anomaly $\epsilon-3p$ v.s $T$ in (d) of FIG.
\ref{EoS}. At $\mu=0$, the sharp peak with a infinite slope edge
indicates a phase transition. With the growing chemical potential $\mu$, the
peak becomes wider with decreasing height. The phase transition turns to be
weaker and weaker for increasing $\mu$ and eventually becomes a crossover.

\setcounter{equation}{0}
\renewcommand{\theequation}{\arabic{section}.\arabic{equation}}%

\section{Conclusion}

In this paper, we studied the Einstein-Maxwell-dilaton system. We obtained a family of analytic black hole solutions by the potential reconstruction method. We then studied the thermodynamic properties of the black hole backgrounds. We computed the free energy to get the phase diagram of the black hole backgrounds. In its dual holographic QCD theory, we are able to realized the Regge trajectory of the vector mass spectrum by fixing the gauge kinetic function. We then discussed the possible interpretations of the phase structure that we obtained from the gravitational background by comparing the lattice QCD simulations. We argued that the heavy quarks interpretation is more favored in our system. We calculated the equations of state in our holographic QCD model. We found that our dynamical model captures many properties in the realistic QCD. The most remarkable feature of our model is that, by changing the chemical potential, we are able to see the conversion from the phase transition to a crossover dynamically. We identified the critical point in our holographic QCD model and calculated its value. As the authors knowledge, our model is the first holographic QCD model which could both dynamically describe the transformation from the phase transition to the crossover by changing the chemical potential and realize the linear Regge trajectory for the meson spectrum.

Since we interpret our holographic QCD model as a heavy quarks system, it would be interesting to perform a lattice simulation on the equations of state of a heavy quarks system and compare with our results. On the other hamd, there are many future directions one can study. For example, the most interesting issue is to find an appropriate warped factor such that one can obtain a phase diagram similar to the common QCD phase diagram. Another interesting issue is to incorporate the chiral symmetry breaking by introducing a scalar coupled to the flavor fields and take the backreactions of flavor fields into account. It is also interesting to compute the linear quark-antiquark potential and expectation value of Polykov loop. One can compute the meson spectrum and determine the quarkonium dissociation temperature in our background. One can also compute the various transport coefficients like shear visocisty, bulk viscosity and so on. It is also interesting to compute the critical exponents of various physical quantities near the critical point. Some of these issues are in progress.

\begin{acknowledgments}
We would like to thank Rong-Gen Cai, Chung-Wen Kao, Prasad Hedge, Mei
Huang, David Lin, Xiao-Ning Wu, Qi-Shu Yan for useful discussions. SH
is also very thankful to the organizers and participants of Symbolic
Computation in Theoretical Physics: Integrability and
super-Yang-Mills held in ICTP-SAIFR, Sao Paulo, Brazil.
SH also would like to appreciate the partial financial support from
China Postdoctoral Science Foundation No. 2012M510562. This work is supported by the National Science Council (NSC 101-2112-M-009-005 and NSC 101-2811-M-009-015) and National Center for Theoretical Science, Taiwan.

\end{acknowledgments}

\end{document}